# Artificial Data, Real Insights: Evaluating Opportunities and Risks of Expanding the Data Ecosystem with Synthetic Data

Richard Timpone[1] and Yongwei Yang[2]

[1]Global Science Organization, Ipsos
[2]Gemini UX, Google

Synthetic Data is not new, but recent advances in Generative AI have raised interest in expanding the research toolbox, creating new opportunities and risks. This article provides a taxonomy of the full breadth of the Synthetic Data domain. We discuss its place in the research ecosystem by linking the advances in computational social science with the idea of the Fourth Paradigm of scientific discovery that integrates the elements of the evolution from empirical to theoretic to computational models. Further, leveraging the framework of *Truth*, *Beauty*, and *Justice*, we discuss how evaluation criteria vary across use cases as the information is used to add value and draw insights. Building a framework to organize different types of synthetic data, we end by describing the opportunities and challenges with detailed examples of using Generative AI to create synthetic quantitative and qualitative datasets and discuss the broader spectrum including synthetic populations, expert systems, survey data replacement, and personabots.

While Synthetic Data has been in the research toolbox for decades, a surge in interest and potential has grown with the advances of generative AI. This article is intended to cut through the confusion and hype of what it is, how it falls in the research data ecosystem, and how it can be evaluated to ensure it is used responsibly, just as researchers have done with other data sources in the past.

*We define synthetic data as computer generated, artificial information as opposed to primary and secondary data captured from real-life individuals or real-world events, which include numeric, categorical, text, image, video and other data forms.*

The breadth of this definition highlights that it is not a single method or approach but rather a broad domain that covers such diverse research needs as fusion and imputation, synthetic respondents and AI agents. In our approach, it broadly encompasses computer

Paper originally prepared for the International Conference for Computational Social Science, Philadelphia, July 2024. The authors would like to thank Christopher Graziul for his comments as well as Joe Paxton, Mario Callegaro, Jim Legg, Ajay Bangia, Maya Ilic, and Dragos Man for their input.

generated, artificial data.  It does not include simple or advanced manipulation of primary data as in approaches like weighting or multilevel regression with poststratification (Gelman and Little 1997; Valliant et. al. 2018).  Specific features and elements of evaluation vary by the nature of the generation and use of specific synthetic data implementations.

We do not necessarily see primary and synthetic data at odds with each other, but rather as part of the broader data ecosystem where each will provide value in different use cases. Furthermore, hybrid or augmented data, combining both, may be leveraged as well. However, to do so responsibly, the idea of humans-in-the-loop (HITL) is critical as researchers play an important role in determining the appropriateness of each type of data for their research use case and the best way to implement the specific methods to generate the needed synthetic data (Daugherty and Wilson 2018).

Our evaluation and consideration of the different elements of the data ecosystem and methods of creating synthetic data also build on the framework of *Truth, Beauty, and Justice* (Lave and March 1993, Taber and Timpone 1996).  Moreover, given the diversity of the use cases of synthetic data, the relative importance of the truth, beauty, and justice dimensions in the evaluation varies as well, with each of the three taking the primary role in different cases.

There are multiple reasons that synthetic data is being embraced so strongly with the new AI methods.  First,  there can be potential gains in terms of greater speed and lower cost of research without having to collect additional primary data. Second, extending the idea beyond simple operational efficiencies -- we may have a reduced need to collect new data, extracting greater value from what is already collected and addressing other logistical challenges.  This is particularly the case to expand empirical testing to areas that may be more challenging to do so.  While representativeness is a risk that will be considered in evaluation, for hard-to-reach groups it is possible that augmenting primary data with synthetic data for these groups could aid representation and ensure these voices are more fully heard in the research (although see Lundberg et al. 2024, Small 2024).

Next, given synthetic data is artificial, concerns of privacy and regulation can also be reduced.  This is especially important in health research and areas that are particularly sensitive, where sharing original datasets is limited and it has been used for years.  This also has potential for research and sharing of datasets to address GDPR and other regulations as well as relaxing the sharing of data in sensitive areas for individuals; providing an alternative path to address the call for privacy-preserving data infrastructures for data sharing (Lazer et al. 2020)

Extending this further leads to the issue of measurement more fundamentally and the idea that this could address issues where direct measures may be less amenable.  In Taber and Timpone (1996), we created a typology of models in the social sciences.  This is reproduced in Figure 1 and has two dimensions, one focused on the level of analysis and the second (dubbed Occam's Dimension) dealing with the level of complexity of the model.  As we move up-and-down the level of analysis, direct measurement of some specific constructs may be extremely challenging, and we will discuss how this may be useful for testing theoretic models or synthesizing data.  Furthermore, we will explain the relevance of the level of analysis to synthetic data, as some approaches attempt to directly synthesize aggregate patterns whereas others first synthesize individual elements (i.e. synthetic individual respondents) and then aggregate those.

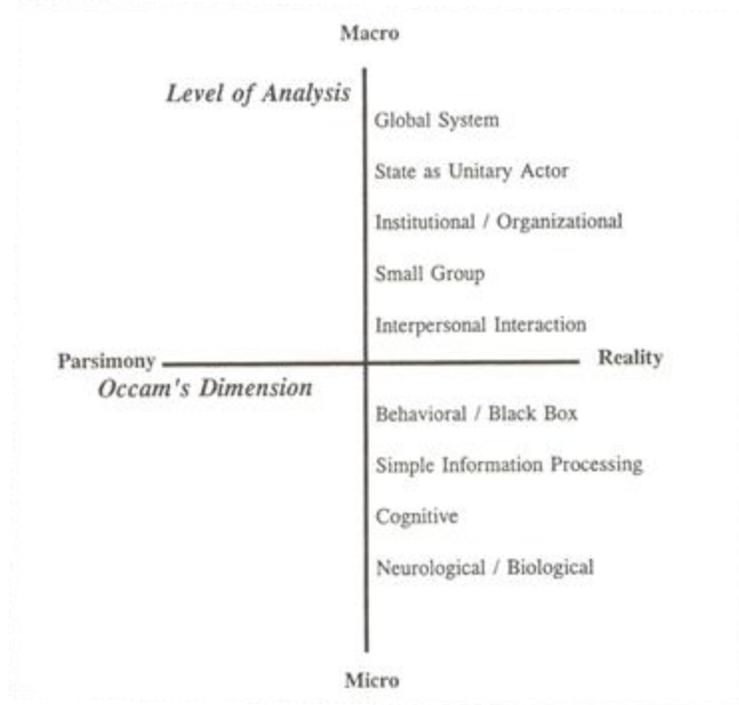

Figure 1. Taber and Timpone (1996) Typology of Social Science Models

For these reasons, we recognize the reality that synthetic data is already part of the data ecosystem, and the future of research will include new forms of it along with hybrid augmented primary+synthetic data in the research ecosystem.  That said, we also see hype being propagated that overstates some of the current potential, leading to some claims that risk being to the detriment of understanding, while others overstate the idea of the death of other parts of the data ecosystem.  Just as with Big Data, the understanding of each source did not lead to the total replacement of other data types as some argued at the time, but rather enriched our toolkits and broadened our testing (Anderson 2008).  In the following sections, we provide a more nuanced discussion of synthetic data and how to evaluate them.

**Synthetic Data and The Data Ecosystem**

We begin by considering the role of synthetic data in the broader data ecosystem, which evolves along with scientific discovery approaches. This places the advances of computational social science in the broader research and data ecosystem (Lazer et al. 2009, Lazer et al. 2020). Jim Gray described three historical paradigms of scientific discovery: *empiricism* (i.e., focused on describing natural phenomena), *theoretical* (i.e., extending to modeling, statistics, and generalizations), and *computational* (i.e., extending to the simulation of complex phenomena) (Gray and Szalay 2007). Moreover, he argued that the techniques and technologies of modern data-intensive scientific discovery are so different from earlier generations that it represents a new *Fourth Paradigm of "data exploration", or eScience*, that unifies theory, experimentation, and simulation (Bell 2009, Hey et al. 2009). This idea of eScience aligns with the foundation of computational social science "that leverages the capacity to collect and analyze data with an unprecedented breadth and depth and scale." (Lazer et al. 2009, p 722).

Most importantly, in the framework of scientific discovery, each past and future paradigm builds on and extends earlier ones. This recognition of integration and continuation is particularly relevant for our discussions about synthetic data, in that we see the ideas behind synthetic data are not invented during the Fourth Paradigm but have weaved through the earlier *empirical*, *theoretical*, and *computational* paradigms (c.f. Hofman et al. 2021). Specifically, we move from the focus of mining Big Data described by the Fourth Paradigm and revisit synthetic data from the perspective of explicitly leveraging algorithmic solutions to produce data for description and modeling.

Empirical evidence remains the foundation of much social science research. Prior to the Scientific Revolution in the 16th century, the collection of such evidence focused on observation, in which modern day quantitative primary and secondary data collection today are rooted (Malik 2017). While scientific discoveries advanced from using simple description to applying experimentation, statistical analysis and modeling, data from real-world human actions remain the main source of empirical data for these discoveries. With synthetic data, however, we are adding a variation of the first, *empirical*, paradigm where human data are being revisited, augmented, and potentially replaced by computer-generated sources for the purposes of description, explanation, and prediction (Hofman et al. 2021).

Similarly, this revisiting of the sources of empirical evidence also plays out in the *theoretical* and *computational* paradigms. In fact, the discussion of artificially generated data is itself a *product* of the tools developed in these two paradigms. On one hand, the idea of synthesizing missing data from observations with statistical imputation and fusion methods has a long history in the social sciences (Little and Rubin 1987). On the other hand, we further note recent research applying synthetic results to statistical model inferences directly (c.f. Li et al. 2023, Moore et al. 2024)

Focusing specifically on the *computational* paradigm, we need to distinguish between algorithmically generated data for testing theoretical representations from data generated to replace real-life data of people or systems. A classic example of data generated to test the implications of theory are the findings of Thomas Schelling that small individual preferences can lead to systematic clustering and segregation (Schelling 1978). More recent examples of simulating psychological and sociological theory are the explorations of the value of diversity (Page 2007; van Veen et al. 2020), and the *John Q. Public* (JQP) model to examine the mechanisms of hot cognition and motivated reasoning (Kim, Taber, and Lodge 2010, Lodge and Taber 2013).  The data predictions and results of JQP were explicitly compared to the observed dynamics of a political campaign and tested against a competing theory of Bayesian updating.

This approach of simulating the effects of theories allows blending the rigor of formalism with more realistic complexity than other formal theory methods (Gilbert and Troitzsch 2005, Horton 2023, Ross 1997, Taber and Timpone 1996, Timpone and Taber 1998).  In contrast, what we are focused on here is the use of analytic and algorithmic techniques to generate valid data points to stand in place of actual observed data.  These range from traditional statistical methods to computational models, Machine Learning methods and new generative AI techniques.  The diversity of methods reflects the breadth of the types of data and use cases we will see in the next section.

Before moving to explore the breadth of types of synthetic data, it is important to clarify that the discussion of adding synthetic data to the ecosystem is not advocating for it to replace other types of data.  As we will emphasize, evaluation of synthetic data for its purpose is critical to ensure its' responsible use.  A clear example of the value of different types of information in the data ecosystem was provided by the mass collaboration exercise in the Fragile Families and Child Wellbeing Study (Salganik et al. 2020).  In spite of over 13,000 variables measured over time for over 4000 families, 160 teams of researchers could not accurately predict a set of life outcomes for the participants with methods ranging from statistical to advanced machine learning and AI.  Conducting 114 in-depth qualitative follow-up interviews provided insights that the data mining could not (Lundberg et al. 2024, Small 2024).  This not only highlights the value of primary empirical data, but also more qualitative methods in the research framework as well. In fact, qualitative insights will continue to prove valuable in helping generate strong theories as we go from data to theory to data, including synthetic data in the broader ecosystem (c.f. Grigoropolou and Small 2022).

It is our view that each pillar of the data ecosystem will continue to provide a key role, even as we believe synthetic data will have a growing place going forward. That said, as evaluation of newer components and methods fall short, others, both quantitative and qualitative, will be key for deriving insights and further improving methods. In the next section we provide a way to organize the diverse approaches to create and understand the nature of the synthetic data outputs before delving further into the questions of evaluation.

**The Breadth and Diversity of Synthetic Data**

Synthetic Data is a broad domain of research and collection of multiple analytic and algorithmic techniques. In this way, it is more like predictive analytics, than a single specific method itself. However, individuals often have a specific use case in mind when discussing synthetic data. This leads to some confusion because different techniques and use cases have different purposes and may require different considerations for quality and evaluation.

The breadth of synthetic data covers approaches as broad as imputation and fusion, artificial respondents and insights, and AI assistants and agents. While synthetic data is not new in social science research, the advancement of generative AI creates new methods as well as fundamentally new use cases for where and how it could replace or augment primary data in the data ecosystem. To help clarify the diversity of types of synthetic data, we leverage a typology built on two dimensions. The first is the nature of the data that is created, from structured quantitative datasets to more qualitative and free-form human-like responses, and the second is the nature of the data generating process, from deterministic and analytic techniques through newer generative AI methods. The framework is provided in Figure 2 below.

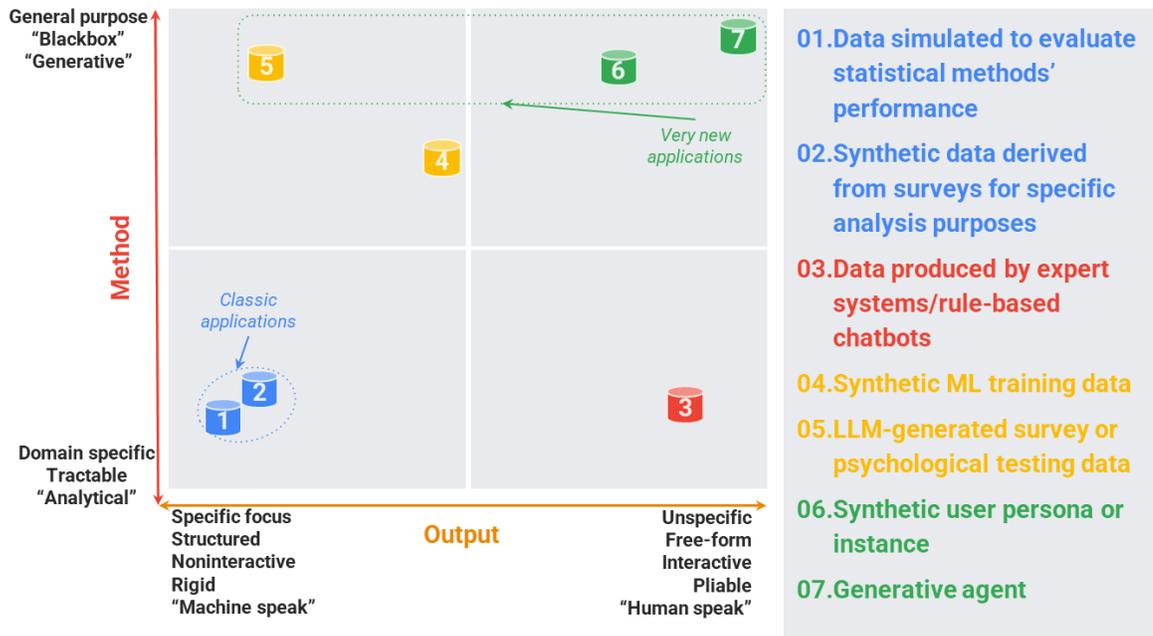

Figure 2- Synthetic Data Typology

While this typology focuses on the data generating processes and the nature of the data, the purpose for which the data is to be used can vary as well. The nature of these use cases also has implications for the evaluation and implementation of synthetic data.

While techniques will continue to evolve, we can describe those used in each quadrant of Figure 2 to provide greater structure for understanding. In the lower left quadrant, we have the more traditional methods of analytically generated structured data. This includes data produced from monte carlo simulations for testing performance of statistical methods (Mooney, 1997). It also includes imputation of missing data, fusion of multiple data sources with traditional statistical methods including synthetic populations (Little and Rubin 1987, Page 2024). Moving up the Y axis and into the upper left quadrant, we encounter more complex machine learning methods such as Generative Adversarial Networks to create structured data directly and through their role in building Large Language Models- LLMs (Langr and Bok 2019).

Before moving to new generative AI methods, we also want to note the lower right quadrant, which includes traditional methods intended to mimic human interactions. In this domain there are algorithmic approaches and computational models going back to the classic *ELIZA* program designed to mimic interaction with a psychotherapist (Weizenbaum 1966). Later, more structured approaches emerged, including expert systems and other knowledge structures that may be useful on their own or incorporated as Knowledge Graphs in LLMs (cf. Benfer et al. 1991, Bratanič and Hane 2024, Taber and Timpone 1994). While many of these traditional methods are still relevant, the advent of generative AI has created new opportunities to synthesize data and create human-like interactions which have accelerated the questions about what is possible with synthetic data for research going forward (Wolfram 2023).

Moving to the very top of the upper quadrants of the typology, we focus on newer generative AI methods, esp. LLMs. These methods have been used to create structured quantitative data in the upper left quadrant (e.g., Argyle et al. 2023, Dominguez-Olmedo et al. 2024, von der Heyde et al. 2024). Techniques on this range from creating tabular datasets from LLMs to synthesizing individual observations for further exploration and analysis. The latter involves first creating "silicon samples" or "AI Twins" to actual respondents. Next, responses generated from them are used for analysis like in human dataset collection. In contrast, aggregated data may be generated directly, bypassing individualized levels of analysis. This returns to the Figure 1 taxonomy as synthetic data may be created at different levels of analysis even for similar uses.

Finally in the upper right-hand quadrant, we have the most recent contributions to synthetic data with Generative AI results mimicking and replacing human interactions. In these cases, they may reflect individuals, or archetypes of segments in the population and play the roles of AI assistants or, at the extreme, AI agents that have role-directed goals and the potential to act autonomously.

**Importance of Evaluation**

Given that the intent of synthetic data is to take the place of human generated primary data, the evaluation of solutions for the different uses of such data is critical to protect scientific integrity (c.f. Abdurahman et al. 2024; Blau et al. 2024). Previously, we had leveraged Lave and March's (1993) criteria for evaluating models in the social sciences to our evaluations of the family of computational and AI models (Taber and Timpone 1996, Timpone and Yang 2018). These criteria are the dimensions of *Truth, Beauty* and *Justice*:

Truth: Degree to which data and models are accurate in what they intend to reflect

Beauty: How data and models address parsimony and the fertility of use cases

Justice: How the production and use of data and models impact human quality of life

In this section, we adopt these three evaluation dimensions to evaluate synthetic data and discuss their interplay.

Truth is often viewed as the most important of the criteria and it has significant implications for the other two dimensions. However, passing thresholds for success in one is not sufficient to avoid issues in the others. Thus, considering the three as separate dimensions of evaluation is necessary.

While seemingly straightforward, unfortunately, we know that there are multiple ways to present results which can lead to misleading findings of accuracy (i.e. Huff 1983). For example, when someone says their results are 80% accurate, the mental model of research is often what is seen in Figure 3 of noise from estimates or synthetic data points around the true values they represent. However, in some cases it may be that the results are the same in 80% of the cases but lead to the exact opposite or erroneous position in 20% of the cases. The idea of being totally misled with a 1 in 5 chance is a much bigger risk than what people think of with the mental model of Figure 3.

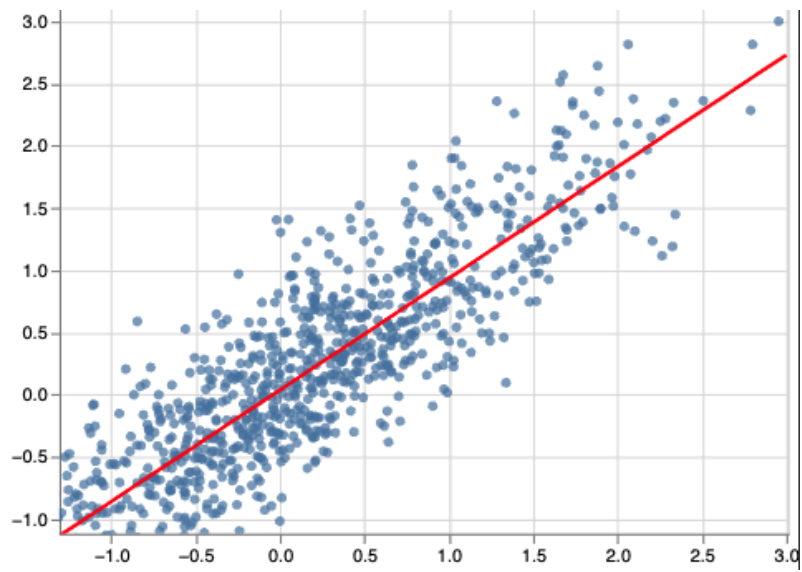

Figure 3. Scatterplot of data with regression line with an $R^2$ of .89 (r=0.8)

This is further illustrated in the aggregate table in Figure 4. Here a specific behavior was exhibited by 6% of the population, and the synthetic data produced by fusion methods also estimated 6% of the population. In this case, the researcher had test and holdout data and it was possible to look at the results conditioning on the actual data. In this case, it was observed that there was not a single observation where the person who actually exhibited the behavior had synthetic values that predicted this. Thus, while the researcher could state that the synthetic data was accurate 88% of the time, if one looked at the substantive cases of who exhibited the behavior or was predicted to do so, there was 100% misprediction! This highlights the critical nature of what the purpose of the data is for. If only univariate summary stats were desired, the data was extremely accurate, if trying to look deeper and examine any data inter-relationships at the unit of analysis, there is substantial danger. Again, this raises the level of analysis question for the validation of data leveraging the framework in Figure 1.

**Synthetic Data**

|  | Yes | No |
|---|---|---|
| **Actual Primary Data** Yes | 0% | 6% |
| No | 6% | 88% |

Figure 4. Confusion Matrix Example

This idea that mental models of problems with accuracy being less serious than they may actually be, extends to other data than structured numeric data types as well. Researchers have found, for instance, that speech to text built off LLMs hallucinate in 1.4% of the cases by injecting totally fabricated sentences when no one was speaking (Koenecke et al. 2024).  Worse, as in the numeric examples, these are not minor issues, but in this case 38% of the hallucinations were actually harmful, with text blocks categorized as perpetuations of violence, dangerous inaccurate associations, or statements of false authority.  In this way, the lack of accuracy leads to further issues regarding the criterion of Justice as well.

Given that data and models are representations of a world we intend to understand and influence, the consideration of beauty focuses on 'aesthetic' and utilitarian aspects.  That is, we consider whether these purposeful representations are appropriately parsimonious while, at the same time, generate many testable hypotheses and surprising results (Gilbert and Troitzsch 2005, Lave and March 1993, Starfield et al. 1990).  As we consider synthetic data, an aspect of this dimension is whether the richness of human and social data is adequately recreated for descriptive and creative uses (see for instance Lodge and Taber 2013, Ross, 1997, Timpone and Taber 1998).  As we will see, this varies in importance based on the purpose of the data itself.  This and the importance of explainability vary across the dimension from the modeling focus on questions of parsimony to how far formalisms should go in modeling processes from simpler analytic to richer algorithmic ones.

"Justice takes into account the real-world implications of a model, especially those that affect quality of life" (Taber and Timpone 1996).  Here we are covering both the externalities of economics and the "consequence" aspect of modern test validity theory (cf. AERA/APA/NCME Standards).  This encompasses multiple important areas – AI ethics, bias, algorithmic fairness, data security and privacy, societal implications including human, work and environmental, alongside the rights and responsibilities of creators of data used for training, and those using it, including the disclosure of the use of synthetic data.

The evaluation of Justice is strongly tied to Truth.  Moreover, given the ubiquity of data as well as model and algorithm-based decisions in peoples' lives today, when we examine the practical implications of the use of data and models, the importance of Justice and ethical considerations would be rising in importance and require researchers to consider this far more actively and thoroughly (Eubanks 2017, O'Neil 2016, Timpone and Yang 2018).

Still, depending on the specific use cases of data, one may weigh each dimension differently for evaluation.  If the goal is the aggregated central tendency, the accuracy of that compared to human response and behavior is key, and focusing on aspects of Truth is appropriate.  However, if we are building respondent level data, the full representativeness and variability in human data would need to be reflected as well (c.f. Atari et al. 2023).  Here and in generating new ideas and hypotheses, the richness and surprise of the domain of Beauty gets elevated.  Finally, in those areas, such as AI agents, where autonomous action is the intent, Justice may predominate.

Moving from the quantitative example to more free-form responses, even the same tool may need different evaluations based on distinct intended uses.  For instance, if a personabot is developed to produce new concepts, ideas, and hypotheses, the richness and fertility of the ideas generated may be more important than accuracy in matching what humans would have produced in the same situation (c.f. Horton 2023).  If the intent were to replicate and replace human input, the accuracy criteria would be the primary one instead.

While the intent of the use case determines the appropriate type of evaluation, other elements matter as well.  For instance, if using Foundation Large Language Models, accuracy changes over time, especially in dynamic markets and their future accuracy may be further at risk by training on generated data.

**Examples of Approaches from Across the Synthetic Data Typology**

The best way to bring to life the diversity of the broad domain of synthetic data is to provide a few examples from across the typology in Figure 2. While users may not care about the specific data generating processes and other elements underlying their use cases, these are key to ensure the methods are appropriate for their intended purpose. Our examples focus on newer generative AI methods in the upper section with case studies that range from quantitative survey to a more qualitative context and, finally, to even less structured interactions.

It should be noted that the first two cases illustrate situations when researchers leverage LLMs to augment and boost existing data, or to fully replace human data collection. They both build respondent level data but the methods and purpose cover different spaces in Figure 2. The first example is closer to the left side of the upper left quadrant of structured data from LLMs. The second case shifts toward the center with a blend of structure and qualitative use and a variation on the data generating process.

**Case Study 1: LLMs to Synthesize Respondent Level Survey Data**

The first case was conducted by Google UX researchers and is described in Paxton and Yang (2024). It explored how well LLMs could simulate human attitudes about technology products compared to a general population survey. The design compared synthetic data produced by prompting GPT 4o and Gemini Pro 1.5 with personas that matched groups aligned with human respondents' age, gender, and self-reported product usage. This approach to generating synthetic data for group-based prompts is comparable to other research in the literature (Bisbee et al. 2024, von der Heyde et al. 2024). The synthetic datasets were then compared to data from 500 human respondents collected in the 1st half of 2024 from Qualtrics Panels. The results specifically focused on comparisons of 8 attitude questions, mostly related to trust and privacy.

This case study was focused on quantitative data synthesis and each question was asked on 5-point Likert scales that was recoded to "top 2 box" for ease in interpretation and modeling comparisons. Since the intent was to evaluate the comparability of LLM-generated synthetic data to human responses, the first dimension of consideration is Truth, i.e., the accuracy of the data. Examinations of the data sources were provided for top-2 box percentages of each question as well as the rank order correlations between human data and the two synthetic data sources.

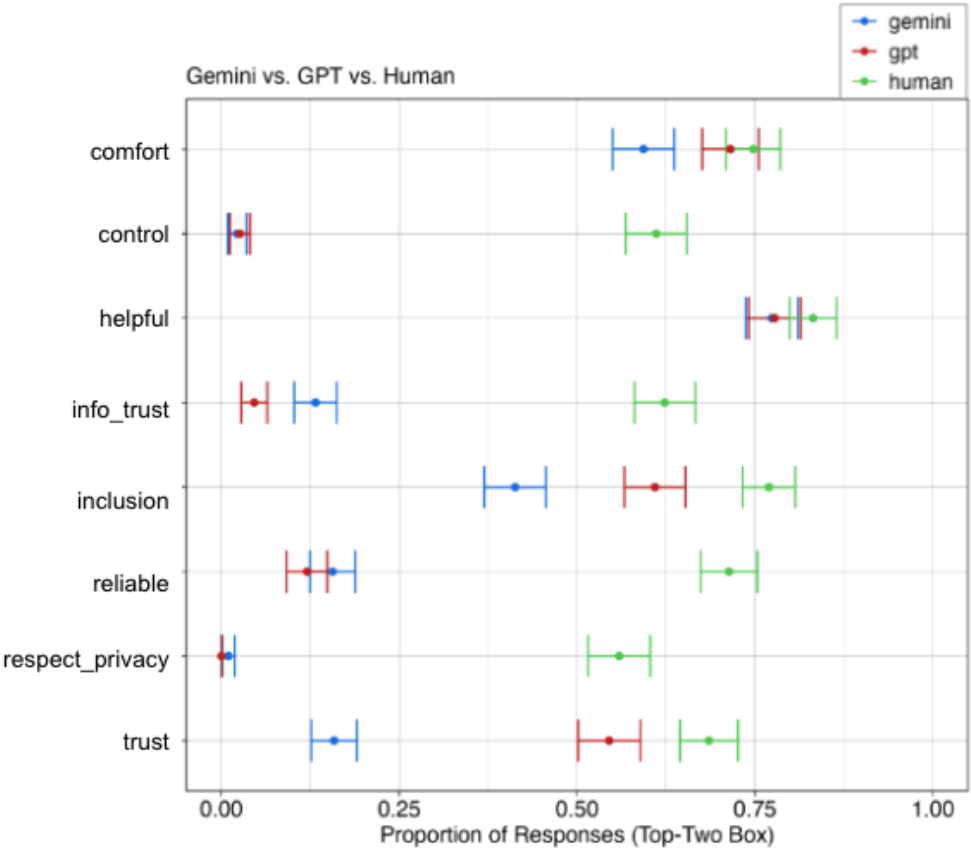

Figure 5. Proportion of Responses (top 2 box) for each data source

The results in Figure 5 show mixed quality in terms of replicating human sample results. While the results are comparable for the helpfulness and comfort of the solutions, there are substantial differences in other questions, especially control, trustworthy information, reliability, and respect of privacy.

Moving to the rank correlations of each question across the 3 data sources, Figure 6 shows a similar story in that a majority of the correlations are low and a large number of them were close to zero. The between-question variations in the size of the correlations are roughly in-line with the pattern observed from mean comparisons. Similar to the models in the mass collaboration described earlier around the Fragile Families Challenge, the two synthetic sources were more related to each other than to the human data they were intended to emulate (Lundberg et al. 2024, Salganik et al. 2020).

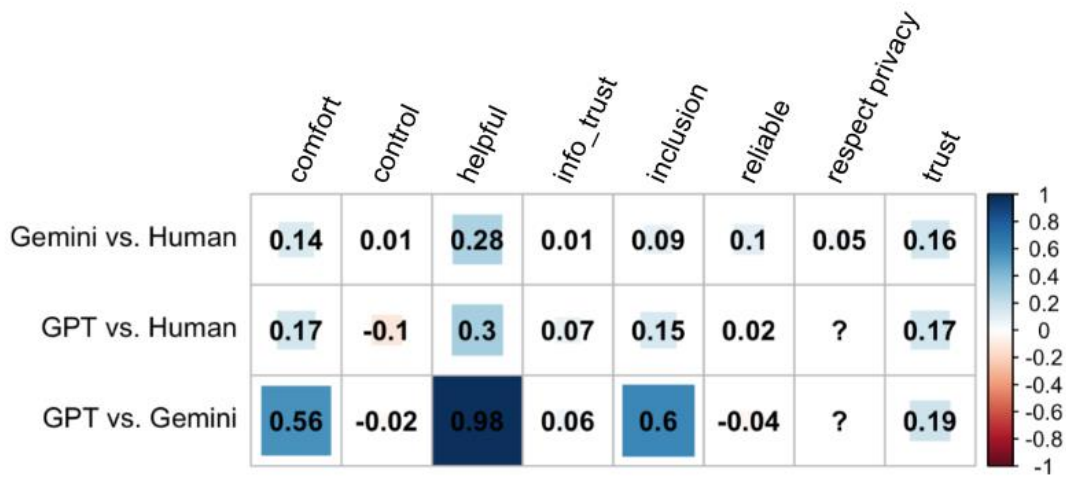

Figure 6. Cross Data Source Rank Correlations
(Note: "?" indicates correlations could not be computed due to zero-variance in the LLM-generated data)

Recent research has found weakness in zero shot estimates, as conducted in this study, for producing synthetic survey data (Abdurahman et al. 2024; Rakovics and Rakovics 2024), and classification tasks (Møller et al. 2024). Even beyond this though, a number of studies have shown inconsistency in LLMs ability to mimic human survey responses more generally (Abdurahman et al. 2024, Bisbee et al. 2024, Dominguez-Olmedo et al. 2024, von der Heyde et al. 2024).  These findings align with the concerns in this type of use and construction of synthetic data, and again highlight the importance of evaluating the quality focusing on accuracy to take the place of human produced data.

**Case Study 2: LLMs for Qualitative Insights**

The second case study provides a variation on the first in terms of data generation and intended use.  The study is described more fully in Bangia et al. (2024).  While still toward the structured side of our synthetic data taxonomy, both the qualitative nature of the training and use represent a horizontal shift toward the center of Figure 2 compared to our first case.  The study was conducted by researchers of Ipsos UU, the company's qualitative research division.  This was produced leveraging an online community of women in Japan who discussed their experiences with, and possible products designed around addressing the physical and emotional effects of menstruation (English translations were used to leverage the deeper quality of the LLMs used).  The creation of the synthetic data shares a resemblance to Argyle et al.'s (2023) 'silicon sample'.  Specifically, separate prompts were created using both qualitative and structured data from training and follow-up activities for each of the 142 participants (all data prompted as text).  Such prompts are dubbed 'AI Twins' and are thus able to explore not only aggregate patterns and inter-relationships in the data but also the fidelity of each AI Twin to replicate the views and sentiments of the community members they are supposed to represent.  Like the example in our first case study, these synthetic AI Twins were created using the best currently available LLMs prompted with individual level past data.  In this case, the synthetic data was produced using Chat GPT 4 32K and Gemini Pro 1.5.

The study tested the quality of the synthetic twins to questions of exploration, evaluation, and ideation. Thus, the concern is with both the Truth and Beauty dimensions, of accuracy and richness of insight generation, while the respect and privacy elements of Justice underlie the entire endeavor. The exploration stage of the research project focused on describing physical and emotional needs around different phases of menstruation, while the evaluation phase considered the evaluation of new product ideas. These two phases of the research differed in how far they deviated from the data used to identify each Twin in the prompts (with innovative product ideas being more unique and new). While showing differences from the training, given the intent of the data, the primary aspect of their evaluation was how accurately the synthetic data reflect the human responses.

Given the closer alignment of the data in the prompts, it is not surprising that the exploration questions showed reasonable ability to identify the main key themes raised by the actual community members, while the evaluation of truly new product ideas led to significant deviations. Figure 7 shows the significant degree of deviation when asked to choose among four unique product ideas, with a substantial difference in the 'winning' concept between the actual women and their AI Twins. While the greater deviation in the new product comparisons that were less related to the previous data included in the prompts is no surprise, even in the attitude explorations there were some differences. The qualitative researchers found, for instance, that the AI Twins were more conceptual than concrete in their examples, less able to transmit the full range of emotions or grasp social factors on behavior, and were less specific when answering more novel questions (Bangia et al. 2024).

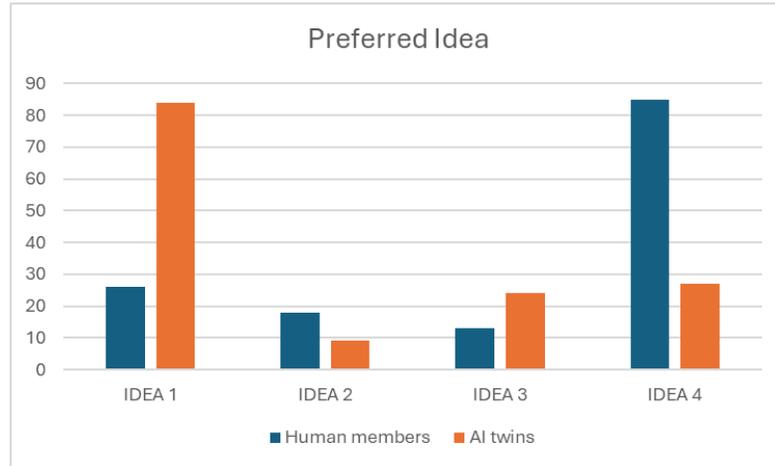

Figure 7. Counts of New Product Idea Preferences of Actual and Synthetic Respondents

Separate from where the synthetic data was able to replicate or deviated from the main themes in attitude exploration and new idea evaluation, both also found patterns of smaller variance in the synthetic data sources that has been seen in a growing number of studies. Lower diversity and variance in LLM data has been seen in synthetic survey data (Aher et al. 2023, Bisbee et al. 2024, Dominguez-Olmedo et al. 2024, Park et al 2023, Sun et al. 2024, von der Heyde et al. 2024), analysis (Moore et al. 2024), text analysis (Abdurahman et al. 2024; Veselovsky et al. 2023), and writing (Padmakumar and He 2024). In the exploration stage of the research, women were asked about physical and emotional needs and while the key themes generally aligned, AI Twins identified fewer than the actual human members as illustrated in Figure 8.

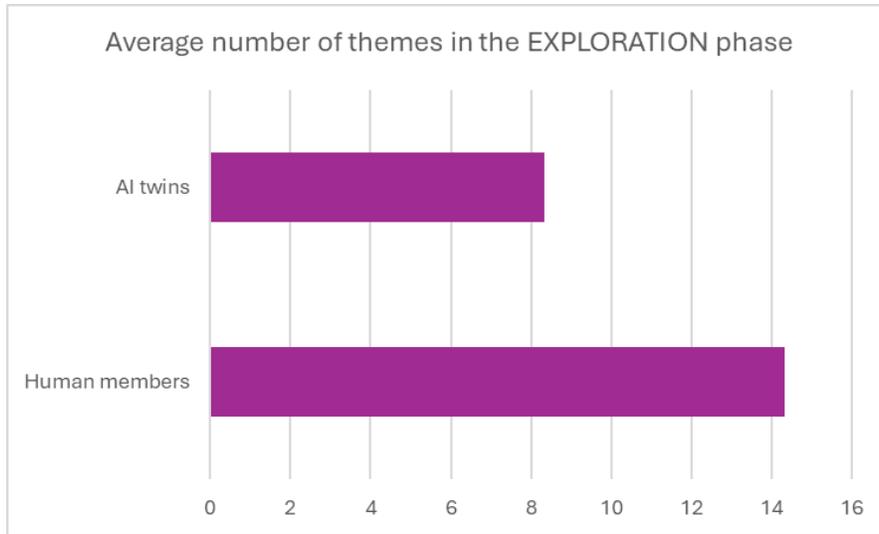

Figure 8. Number of physical and emotional needs identified by each sample

While different in the accuracy of replicating themes, similar to the attitude exploration, Figure 9 shows the lower dispersion of themes identified by the AI Twins around new product ideas as well. While these results match many studies in the literature, new advances in leveraging masses of diverse personas will be worth exploring if they can improve these patterns in the future (c.f. Chan et al. 2024).

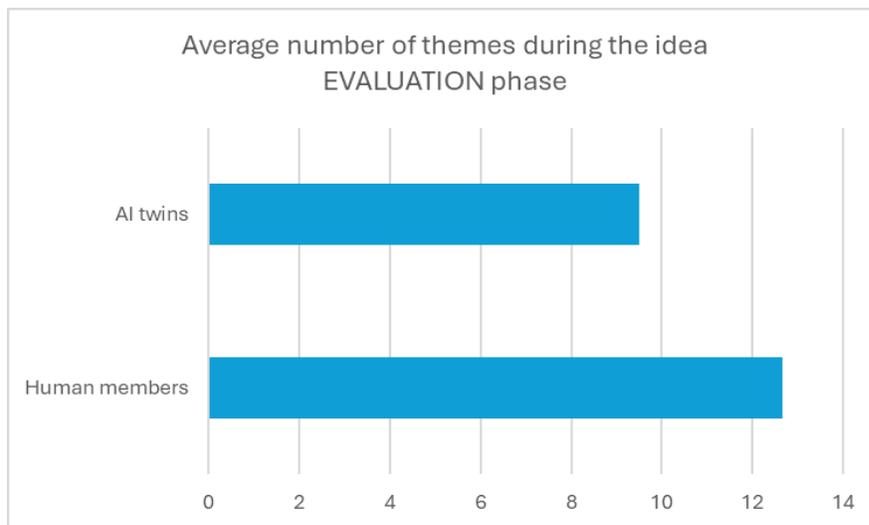

Figure 9. Number of themes identified in new product evaluations by each sample

Note that the lower dispersion and variance identified in the literature is a significant concern as this is leveraging the best current LLMs that leverage data that is likely cleaner than it will be in the future.  While it goes by multiple names (model or mode collapse, model autophagy disorder), researchers have shown that training models on generated synthetic data leads to further reduction in variance and/or quality (Alemohammad et al. 2023, Langr and Bok 2019, Shumailov et al. 2023).  Thus, as more text, images and other content created by generative AI tools are added to the internet, foundation models trained off these sources will continue to deteriorate in quality and variance, further raising the need for evaluation and augmented samples.

Thus far, with this case study, we have focused on the areas where the primary question was the ability of synthetic data to replicate human responses.  The third domain of the investigation though was in the area of creativity and ideation.  There are two equally valid ways to consider ideation.  The first way is to consider whether the ideas generated by synthetic data match those of humans in the same area.  This approach aligns with the standard focus on Truth, esp. accuracy.  The other way is about the idea of fertility of models and data, which falls into the Beauty dimension.  In this case, one could focus on whether the synthetic data is producing new hypotheses and ideas that are valuable, regardless of whether they align with what humans would create in the same case.  Here, it is not if machine intelligence matches human intelligence, but whether it provides value in itself (c.f. Horton 2023, Taber and Timpone 1996).

In this case, the research team felt that the AI Twins provided fruitful value, even though differing from the human respondents.  As the paper by Bangia et al. (2024) shows, while the synthetic data ideas focused less on emotional benefits, they often were more holistic with stronger reasons to believe and linking the product ideas to benefits.  In the domain of ideation, one could weigh creativity and fertility of the results more heavily to evaluate the 'quality' of the data, and using this criterion, the models do perform better. An example of the full richness of the creativity of new product ideas by the AI Twins is provided in the appendix. As can be seen there, these are extremely detailed ideas with elements (like naming them) that human respondents did not generally provide. This aligns with other work in the literature that, while not necessarily better, AI generated ideas do differ from those from humans and can increase total collective idea diversity when combining AI ideas with human brainstorming (Ashkinaze et al. 2024).

While there are debates around the nature and extent of the true creativity of AI and LLMs, these findings demonstrate that these tools can provide value and augment efforts around scientific hypothesis generation as well as practical problem solving and product creation. While AI has already served to enhance and understand user and free, as opposed to producer, product innovation, this demonstrates the potential to go further in the ideation process as well (von Hippel 2017, von Hippel and Kaulartz 2021). Thus, along with the exploration of needs, producer and user generated ideas for identifying potential products and collective problem solving from business to public policy, new Gen AI tools may augment and increase the diversity of human-AI ideation in valuable ways especially if they differ from what people would identify directly, and not in spite of this (c.f. Peach et al. 2021).

**Considerations Beyond the Two Case Studies**

As we explore the full Synthetic Data Typology further, it is worth first staying in the upper left quadrant but demonstrating with different data a similar pattern we have found, and is seen elsewhere in the literature, of less coverage produced by synthetic data than that collected by human respondents. As our definition of synthetic data stated, it may be quantitative, categorical, text, image, etc., and from the typology of Social Science models applied to data from Figure 1 as well, we saw that it can vary across different levels of analysis.

One domain of methods that has been used in research to identify where people's attention is focused is eye tracking methodology.  While we will not explore the various approaches and their validation, this field spans the use of calibrated glasses to track gaze to leveraging cameras on computers to broader approaches that are designed to track individuals and groups with other video tools.  Recently, some AI companies have been training models and claim to be able to create synthetic alternatives to measuring groups of individuals to simulate where people's attention on images and videos would be focused.

This eye tracking example is a clear case of the question of use and the consideration of the levels of analysis. While in the case studies, we simulated respondent level data at the same level of analysis as primary data collection, in these other cases, what is produced is a simulation of group focus directly instead of multiple individuals being aggregated.  Thus, in these cases evaluation compares simulation of group data against actual individuals whose responses are aggregated to the comparable level of analysis.

Figure 10 demonstrates where two synthetic data eye tracking solutions estimate the attention would be focused compared to the aggregate results of a sample of people on a static image (from Pinterest influencer, Stefanie Taylor; https://www.pinterest.com.mx/pin/39054721766874698/).  Here we can see that the two models perform slightly differently in identifying the main areas of attention, but as we saw in the quantitative and text based qualitative findings, the AI models demonstrate less dispersion and variance than that covered by actual people in each.

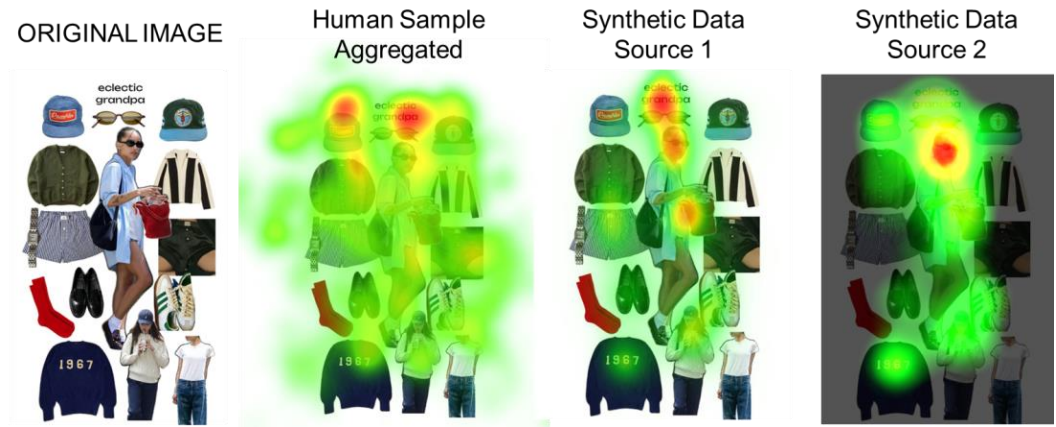

Figure 10. Eye Tracking Variations Across Different Synthetic Data and Actual Tests

Continuing our journey around the Synthetic Data Typology, we move to the upper right quadrant of Figure 2. Here we continue to explore data that is created leveraging newer generative AI tools and techniques but move to more unstructured interactions and qualitative engagements. As described in the typology section, this includes personabots (i.e. chatbots trained to emulate specific personas, whether a type of person, or a segment in a population). This overlaps with the second case study where a prompt directed an LLM to behave like a specific individual to produce responses to structured queries. As we move further to the right on the output or use dimension of the typology, we can engage the personabot directly in more general and less structured discussions.

Just as we saw in the second case, the purpose for which the personabot is intended will create different foci for their evaluation. If it is intended to represent an individual or group, evaluation of its responses will need to be focused on how representative and accurate they are. Just as we saw in the second case study, the further we are from the training data, the less accurate the responses are likely to be.

However, as in case 2, if the intent is to generate new ideas and hypotheses about the segment, then focusing on the criteria of Beauty again and its ability to surprise and produce fertile ideas for exploration is a more appropriate criteria for evaluation.  Thus, the same tool, using the same type of interaction, may again require different criteria for evaluation for the different intended purposes.

Moving further to the right of the upper quadrant of the typology we shift to AI assistants and AI agents.  AI assistants could be evaluated by either accuracy in representing a human assistant in the same way or by its helpfulness from a rich, beauty, perspective instead.  However, as we move further right, we feel that the criteria may shift again.  With AI Agents, we are focused on goal-driven tools that can take autonomous action.  The action may be to make recommendations or to take some type of physical action.  Thus, these are the generative AI extensions of tools that AI models have been improving on in areas from judicial and welfare recommendations to autonomous software controlling self-driving cars.  As we discussed in our earlier work, as we move into these spaces, the criterion of Justice becomes central (Timpone and Yang 2018).

We want to complete discussions across the full typology and not only focus on new Generative AI methods.  Thus, we will briefly discuss the more traditional areas in the lower two quadrants of the synthetic data typology.   The lower left quadrant of Figure 2 is the area best known to researchers, as it is the domain that has been used and tested extensively for years.  This is where analytic and machine learning methods are used to create structured synthetic data sets. One application already mentioned is the creation of Synthetic Populations.  This method results in datasets as if everyone in a census provided all of the demographic, behavioral and attitudinal information desired (Page 2024).  It often leverages fusion and other methods across multiple inputs.  The accuracy of these datasets is primary, and testing varies based on whether one is looking for aggregate descriptives or multivariate insights.  Evaluation also varies if central tendencies or if the diversity across individuals for examining sub-groups or multivariate relationships like the drivers of behavior is the intent.

The lower right quadrant reflects less structured human feedback although still using more traditional analytic methods.  An example here is Expert Systems that are designed to reflect human experts' advice to evaluate and solve problems and are related to chatbots that mimic expert interviewers by probing on topics.  While accuracy is key, the ability to mimic human interaction is also important.  Moving from deterministic rule-based systems to probabilistic elements improves the richness and accuracy, but these tools are often restrictive in domains and limited in the breadth of their accuracy.  As in the upper right-hand quadrant, the intent is key and whether emulating humans or providing value and new hypotheses, the purpose affects the criteria for evaluation.  Just as in the upper right quadrant, these tools, whether built off rules, statistics, or machine learning, if they are intended to take action, Justice and ethical elements again move to the forefront.

**Conclusion**

The first key point of our article is that synthetic data is a broad research domain and what success looks like and how we need to evaluate each method and data use is not the same.  By leveraging the framework of Truth, Beauty, and Justice and taking a tour around our typology, we are hoping that the value of the theory and frameworks is clearer.  This is also true in positioning the different types of Synthetic Data in the broader research ecosystem.  Here we leverage the tools of eScience and the fourth paradigm to create synthetic versions of empirical observations, modeled and simulated data.  From a computational social science perspective this is relevant from multiple perspectives: 1) the tools used to generate the synthetic samples, 2) synthetic versions of empirical data, and 3) extensions of the computational formalisms intended to explore and test theories.

The use cases provided deeper insights into some of the limitations of using synthetic data to replace primary data, as well as opportunities where it can provide value and ways to evaluate and continue testing, as the tools and capabilities evolve further. We also highlighted that while the tools may get better in the future, we need to be wary as the quality of the data may become a larger concern. This is especially the case given the amount of generated text and images being posted on the internet each day. It is not only non-representativeness and misleading content that we need to be wary of, as training future models on generated data is known to create additional issues on quality and further reductions in variance.

Having a framework for differentiating methods and tools, and criteria for evaluating them will help create common approaches for discussions so researchers are on the same page. In this way, we hope that we can all take advantage of the opportunities with synthetic data appropriately and minimize the risks associated with it.

**Appendix: Actual Example of New Product Idea Produced by an AI Twin**

"My magic wand would create a product that is a blend of a food supplement and a comfort food - let's call it the 'Mood Munchies'.

• 'Mood Munchies' would be a variety pack of small, bite-sized snacks that are easy to consume at any time of the day. Each pack would contain three different types of snacks, each tailored for the three different stages of the menstrual cycle - PMS, periods, and post-period.

• The PMS snack would contain ingredients like Evening Primrose Oil and Chasteberry, which are known to alleviate PMS symptoms. The snack for during the period would be iron-rich, to make up for the iron loss during this phase, with ingredients like spinach and beetroot. The post-period snack would be energy-dense, with ingredients like oats and banana, to help rejuvenate and restore energy levels.

• The taste would vary for each snack, catering to the common cravings during each phase. The PMS snack would be slightly salty, the period snack would have a sweet and tangy taste, and the post-period snack would be sweet.

• The texture would be soft and chewy, making it a comfort food that's easy to consume. The individual packaging would make it convenient to carry around and consume as needed.

• Ideally, one would consume one snack per day during the corresponding phase of their menstrual cycle. However, given that the snacks are made of natural ingredients and are nutrient-rich, they can be consumed more frequently if desired.

• 'Mood Munchies' would be for anyone who experiences menstrual cycles and the discomforts associated with them. It's for the person who wants a natural, tasty way to manage their menstrual symptoms and maintain their nutritional intake during this time.

• What sets 'Mood Munchies' apart from other products is that it's not just a supplement, it's a snack. It doesn't just aid in managing menstrual symptoms, but also caters to the cravings that often come with these phases. Plus, the variety pack concept allows for tailored nutrition for each phase of the cycle."